\documentclass{ws-mpla}

\usepackage{epsfig}
\usepackage{amssymb,amsmath,amsfonts}
\usepackage{multirow}
\usepackage{rotating}
\usepackage{graphicx}
\usepackage{lscape}
\usepackage{hyperref}

\begin{document}

\markboth{Hieu Minh Tran, Tadashi Kon and Yoshimasa Kurihara}
{Discrimination of SUSY breaking models at linear colliders}

\catchline{}{}{}{}{}

\title{Discrimination of SUSY breaking models using single-photon processes at future $e^{+}e^{-}$ linear colliders}

\author{\footnotesize HIEU MINH TRAN}

\address{
Hanoi University of Science and Technology, 1 Dai Co Viet Road, Hanoi, Vietnam\\
Hanoi University of Science - VNU, 334 Nguyen Trai Road, Hanoi, Vietnam\\
hieutm-iep@mail.hut.edu.vn}

\author{TADASHI KON}

\address{
Seikei University, Musashino, Tokyo 180-8633, Japan\\
kon@st.seikei.ac.jp}

\author{YOSHIMASA KURIHARA}

\address{
High Energy Accelerator Reseach Organization, Oho 1-1, Tsukuba, Ibaraki 305-0801, Japan\\
yoshimasa.kurihara@kek.jp}

\maketitle


\begin{abstract}
We examine the single-photon processes in the frame work of supersymmetric models at future $e^+ e^-$ linear colliders. According to the recent experimental achievement, the optimistic polarization degrees for both electron and positron beams are taken into account to enhance the signal-to-noise ratio revealing the observable difference between supersymmetry breaking models. The minimal supergravity model and the minimal $SU(5)$ grand unified model in gaugino mediation have been examined as examples. We see that after several years of accummulating data, the difference of the number of single-photon events between the two models received from the collider would be in excess of three times the statistical error, providing us the possibility to probe which model would be realized in nature. The result is well suitable for the future running of the International Linear Collider.

\keywords{Supersymmetry; single-photon; linear collider.}
\end{abstract}

\ccode{PACS Nos.: 11.30.Pb, 12.60.Jv, 13.66.Hk, 14.80.Ly}

\section{Introduction}

Supersymmetry (SUSY) has been attracting lots of interests since it gives us a solution to the gauge hierarchy problem in the standard model (SM). Furthermore, the simplest supersymmetric extension of the SM, the minimal supersymmetric standard model (MSSM), predicts a natural unification of gauge couplings at the scale $M_G \simeq 2 \times 10^{16}$ GeV providing a hint about a grand unification theory (GUT). In SUSY models, there exists a supersymmetric partner corresponding to each SM particle.
However, if SUSY is an exact symmetry, it predicts the same masses for the SM particles and their superpartners which have never been observed. So SUSY must be broken in such a way that preserves the property of quadratic divergence cancellation.
To do so, the soft SUSY breaking terms were introduced in the Lagrangian which include gaugino masses, sfermion masses and trilinear coupling constants.

Experimental data shows an important feature that nature is almost flavour independent and CP invariant.  These requirements severely restrict the allowed values of soft parameters in such a way that they insert only tiny flavour changing neutral currents (FCNCs) and small CP phases.
To understand the origin of the soft terms, many SUSY breaking models have been proposed using the technique of spontaneous symmetry breakdown. The common idea of those models is separating the field content of the model into two different sectors. The visible sector contains the MSSM chiral supermultiplets and the hidden one contains the SUSY breaking source. The difference between models lies on the mechanism used to communicate one sector to an other. To avoid the FCNC problem, the interaction between the two sectors needs to be flavour-blind.
Different mediation scenarios lead to different boundary conditions at the extremely high energy scale. Then they in turn result in different mass spectra at low energies giving distinctive signals at colliders. Previously, the mass spectrum has been used as a probe for SUSY models and seesaw mechanisms.\cite{SUSY_probe,GUTs_discrimination,seesaw_probe} Here we consider two typical SUSY breaking models as examples: the minimal supergravity model (mSUGRA) and the minimal $SU(5)$ grand unified model in gaugino mediation (GinoSU5).\cite{mSUGRA1}$^-$\cite{su5-4} In these models, the FCNCs are suppressed by the flavour independent interaction mediating between the two sectors, namely, the gravitational interaction in the mSUGRA and the gauge interaction in the GinoSU5.

In this paper, we study the collider phenomenology of the above models regarding to the single-photon processes at future $e^+ e^-$ colliders, especially the International Linear Collider (ILC). 
The single-photon process is one of the simplest channels in which only one photon goes out of the interaction point, giving the visible energy, and all other particles contribute to the missing energy. Assuming the R-parity conservation, the lightest supersymmetric particle (LSP), which is neutralino in usual SUSY models, is a stable and weakly interacting one. So the invisible final products of the single-photon events are the neutrinos and the lightest neutralino.
The single-photon events have been explored in detail to search for new physics at the PEP (Position Electron Project) and PETRA (Positron Elektron Tandem Ring Anlage) experiments, the TRISTAN (Transposable Ring Intersecting Storage Accelerator in Nippon) experiment, the Large Electron Positron (LEP) Collider and also in the preparation for the incoming ILC.\cite{single-photon_PEP-PETRA1}$^-$\cite{single-photon_ILC9} 
The lower limits of the sparticle masses established by experiments tell that the sparticles must be heavier than their SM partners. It follows that the SUSY signal would be small compared to the SM background since the masses of intermediate sparticles appear in the denominators of their propagator and the integrating region in the phase space is narrower. Hence, the difference between the SUSY signals of models is even much smaller compared to the background.
Thanks to the high center of mass energy and luminosity, the clean environment and the well-defined initial states of future $e^+e^-$ colliders, like the ILC, the measurement accuracies there become very high.
With all of these advantages, we investigate here the possibility to discriminate SUSY breaking models and point out that this type of data can be used to build up an independent constraint on the parameter space.

Starting from the given benchmark points of the parameter spaces which produce a common base for the two models and satisfy various phenomenological constraints,
we present a systematic approach to the single-photon signal in the ILC at the center of mass energy $\sqrt{s} = 1$ TeV, which can be used for the arbitrary polarization degrees of both the electron and positron beams. With the recent achievement in producing polarized beams (see Refs. \refcite{polarization1,polarization2}), given an expected value of luminosity $L = 1000 \; \rm fb^{-1}/year$, we estimate how long it would take to accumulate data such that the difference between the numbers of evens of the two models is large enough to test the models.
This paper is organized as follows: in Section 2, we review the basic ideas of the mSUGRA and GinoSU5 models together with their input parameters at the high energy scale. In Section 3, we present the calculation method and analyse how to suppress the SM background. Section 4 is devoted for the numerical results. Finally, we conclude and give some discussions in Section 5.

\section{Basis of selected models}

The mSUGRA model actually bases on the idea of gravity mediated SUSY breaking in which the hidden sector connects with the MSSM sector through the gravitational interaction.\cite{mSUGRA1}$^-$\cite{mSUGRA11} In this scenario, the supergravity multiplet acts as a messenger to carry the SUSY breaking from the source to the visible sector resulting in the soft SUSY breaking terms of the effective Lagrangian. Inspired by the grand unification at the GUT scale $M_G$, the universalities of gaugino masses, scalar soft masses and trilinear couplings at $M_G$ are assumed in this model. So the number of free parameters here reduces to only four plus a sign making the model very predictive:
\begin{equation}
m_{1/2}, \; m_0, \; A_0, \; \tan \beta, \; {\rm sign}(\mu),
\end{equation}
which are the common gaugino mass, the scalar soft mass and the trilinear coupling at $M_G$, the ratio of the vacuum expectation values of the two Higgs doublets, and the sign of the supersymmetric Higgs mass respectively.

Beside the gravity mediation, one can use an other flavour-blind interaction such as the gauge interaction to mediate between the two sectors. The GinoSU5 model considered here bases on the gaugino mediated SUSY breaking scenario.\cite{gmsb1,gmsb2} In this scenario, the 5-dimensional space-time setup is introduced to separate the SUSY breaking source and the MSSM matter fields. These two sectors reside in two $(3+1)$-branes locating at different fixed points of the fifth dimension which is compactified on a $S^1/Z_2$ orbifold. The gauge supermultiplets live in the bulk and so directly couple to the fields in both branes, giving masses for gauginos at the tree level. Since there is no direct contact between the MSSM matter fields and the SUSY breaking source, the scalar soft masses and trilinear couplings are suppressed at the compactification scale $M_c$. At the low energy region, they are generated from the renormalization group (RG) evolution.

In order to obtain the neutralino-LSP in the gaugino mediation scenario, the compactification scale should be higher than the GUT scale leading to the necessity of embedding our theory into a SUSY GUT.\cite{GUTs_discrimination,spectrum-gmsb1,spectrum-gmsb2} In our study, the $SU(5)$ is chosen to be the grand unified gauge group. The particle content of the minimal $SU(5)$ GUT model is organized as follows: $D^c_i$ and $L_i$ realize the $\bar{\bf 5}_i$ representation, while $Q_i$, $U^c_i$ and $E^c_i$ realize the ${\bf 10}_i$ representation. The $\bar{\bf 5}_H$ and ${\bf 5}_H$ contain the two Higgs doublets needed to break the electroweak symmetry. The other Higgs fields necessary for the grand unification breaking realize the ${\bf 24}_H$ representation of the $SU(5)$ group.\cite{su5-1}$^-$\cite{su5-4} In the GinoSU5 model, the number of free parameters is only three plus a sign:
\begin{equation}
m_{1/2}, \; M_c, \; \tan \beta, \; {\rm sign}(\mu),
\end{equation}
where $m_{1/2}$ is still the common gaugino mass at the GUT scale and $M_c$ is the compactification scale.

\section{Calculation method and analysis}

When comparing the two models, we need to fix a common base for them. Since the final products of the single-photon processes include only one photon and the missing energy carried by the neutrinos and/or the lightest neutralino, we intuitively choose the lightest neutralino mass as a common base for the two models. The mass of the lightest neutralino mostly originates from the $U(1)$ gaugino mass, so by using the same input parameter for gaugino mass at $M_G$ our two models will have the same neutralino-LSP mass.

In our analysis, we always choose ${\rm sign}(\mu) = +1$ and consider the following benchmark points in the parameter space:
\begin{equation}
m_{1/2} = 400 \; {\rm GeV}, \; m_0 = 100 \; {\rm GeV}, \; A_0 = 100 \; {\rm GeV}, \; \tan \beta = 10
\end{equation}
for the mSUGRA model, and
\begin{equation}
m_{1/2} = 400 \; {\rm GeV}, \; M_c = 10^{18} \; {\rm GeV}, \; \tan \beta = 10
\end{equation}
for the GinoSU5 model.

\begin{figure}
\begin{center}
\includegraphics[scale=0.61]{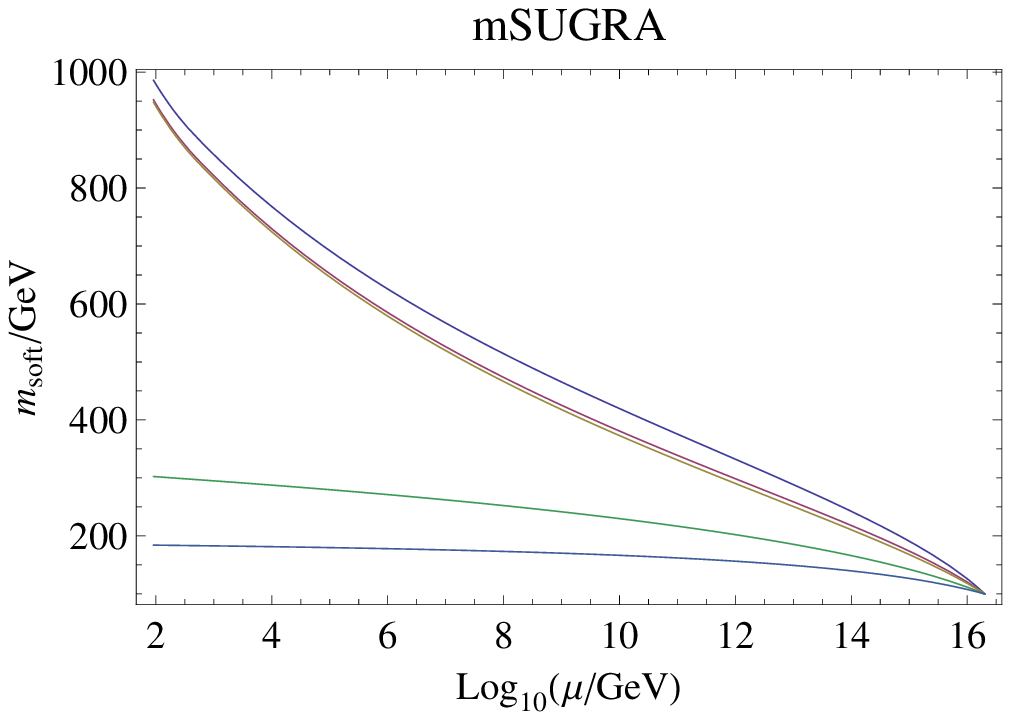}
\includegraphics[scale=0.61]{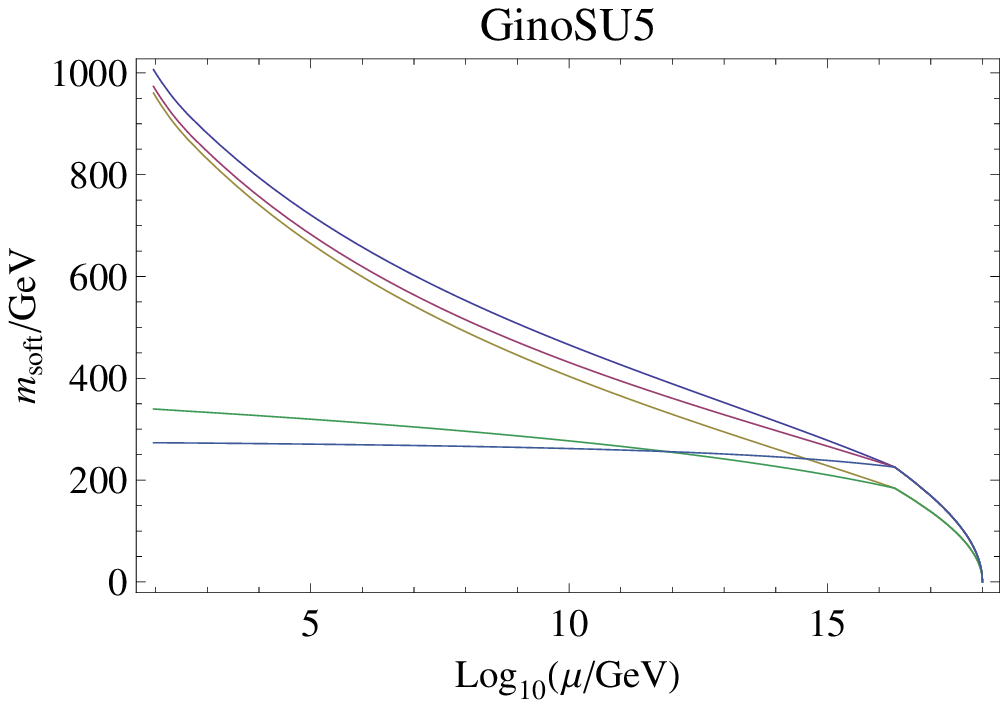}
\caption{Soft mass RG evolutions of the first generation in the mSUGRA model and the GinoSU5 model with the input parameter choices as in the text. In each plot, from bottom to top, the lines correspond to $m_{\tilde{E}^c}$, $m_{\tilde{L}}$, $m_{\tilde{D}^c}$, $m_{\tilde{U}^c}$ and $m_{\tilde{Q}}$ respectively.}
\label{RGevolution}
\end{center}
\end{figure}

To generate the mass spectrum, in the case of the mSUGRA model, we input the universal gaugino mass, the scalar soft mass and the trilinear coupling at $M_G$, then solve the 1-loop MSSM RG equations (Ref. \refcite{rge}) from the GUT scale to the electroweak scale. 
In the case of the GinoSU5 model, after solving the RG equations of the $SU(5)$ SUSY GUT model from the compactification scale to the GUT scale, the values of the soft terms are determined at $M_G$ as follows:\cite{GUTs_discrimination,spectrum-gmsb1,spectrum-gmsb2}
\begin{eqnarray} 
 && m^2_{\bf 10}(M_G) = \frac{12}{5} m_{1/2}^2
   \left[ 1- \left( \frac{\alpha(M_c)}{\alpha(M_G)} \right)^2 \right], \\
 && m^2_{\bar{\bf 5}}(M_G) = m^2_{\bf 5}(M_G) 
   = \frac{8}{5} m_{1/2}^2 \left[ 
    1- \left( \frac{\alpha(M_c)}{\alpha(M_G)} \right)^2 \right], \\
 && A_u(M_G) = -\frac{32}{5} m_{1/2}
    \left[ 1- \left( \frac{\alpha(M_c)}{\alpha(M_G)} \right) \right], \\ 
 && A_d(M_G) = -\frac{28}{5} m_{1/2}
    \left[ 1- \left( \frac{\alpha(M_c)}{\alpha(M_G)} \right) \right],  
\end{eqnarray}
where $\alpha$ is the GUT gauge coupling and
\begin{eqnarray}
 &&  \alpha(M_c)^{-1} =  \alpha(M_G)^{-1} - \frac{3}{2 \pi} 
     \ln(M_G/M_c).
\end{eqnarray}
Subsequently, we solve the MSSM RG equations from the GUT scale to the electroweak scale with the soft term inputs at $M_G$. 
The RG evolutions of the two models for the soft masses of the first generation are demonstrated in Fig. \ref{RGevolution}. We can see that due to the running effect above the GUT scale the soft masses in the GinoSU5 model are heavier than those in the mSUGRA model, especially in the slepton sector.
In both cases, after getting the solutions of the RG equations for the soft SUSY breaking terms, the mass spectra and the mixing angles of the two models are determined from the low energy values of such terms and the experimental data of the SM particles.

With these above choices of input parameters, the two models satisfy the constraint on the Higgs mass lower bound from the LEP 2 data:\cite{Higgs}
\begin{equation}
m_h \geq 114.4 \; \rm GeV.
\end{equation}
Using the micrOMEGAs 2.4 package (Refs. \refcite{micrOMEGAs1,micrOMEGAs2,micrOMEGAs3}), we have checked that the other phenomenological constraints on the branching ratios of $b \rightarrow s \gamma$, $B_s \rightarrow \mu^+ \mu^-$ and the muon anomalous magnetic moment $\Delta a_{\mu} = g_{\mu} - 2$ are also satisfied:\cite{bsgamma,bsmumu,delta}
\begin{eqnarray}
& 2.85 \times 10^{-4} \leq BR(b \rightarrow s + \gamma) \leq 4.24 \times 10^{-4} \; (2 \sigma), \\
& BR(B_s \rightarrow \mu^+ \mu^-) < 5.8 \times 10^{-8}, \\
& 3.4 \times 10^{-10} \leq \Delta a_{\mu} \leq 55.6 \times 10^{-10} \; (3 \sigma).
\end{eqnarray}

Next, the generated mass spectra and the mixing angles are integrated into GRACE/SUSY v2.2.1 in a compatible way.\cite{grace} This package is employed to calculate the cross-sections and the decay widths relevant to our study at the tree level. For a given process, it automatically generates all the possible Feyman diagrams, then produces a FORTRAN source code suitable for further calculation. The numerical integration is performed by the program BASES using the Monte Carlo method. In the output of this step, we obtain the total cross-section together with the differential cross-sections of the process.

Regarding to the single-photon signal, we consider both the SUSY signal and SM background processes. Since only the photon is detectable, the missing energy must be deposited in stable, neutral and weakly interacting particles which in the MSSM are usually the neutrinos and the lightest neutralino. Here, we limit our study to an approximation in which the most significant SUSY contributions to the single-photon signal emerge from the following processes:
\begin{eqnarray}
  && e^+ + e^- \rightarrow \gamma + \tilde{\chi}^0_1 + \tilde{\chi}^0_1, \label{signal1} \\
  && e^+ + e^- \rightarrow \gamma + \tilde{\nu}_l + \tilde{\nu}_l^*,   \label{signal2} 
   \quad l = e, \mu, \tau.
\end{eqnarray}
Since $\tilde{\nu}_l$ and $\tilde{\nu}_l^*$ are not stable, they will quickly decay into lighter particles via the visible channels:
\begin{eqnarray}
  & \tilde{\nu}_l \rightarrow l^- + \tilde{\chi}^+_1, \quad
    \tilde{\nu}_l^* \rightarrow l^+ + \tilde{\chi}^-_1, \\
  & l = e, \mu, \tau.  \nonumber
\end{eqnarray}
and the invisible decay channels:
\begin{eqnarray}
  & \tilde{\nu}_l \rightarrow \nu_l + \tilde{\chi}^0_1, \quad
    \tilde{\nu}_l^* \rightarrow \bar{\nu}_l + \tilde{\chi}^0_1,  \\
  & l = e, \mu, \tau.   \nonumber
\end{eqnarray}
The particles of the visible decay channels leave their tracks in the detector, so only the invisible decay channels account for the single-photon signal.

In general, the signal of new physics often has to face the corresponding huge background from the SM. In our case, the background processes for the single-photon signal are:
\begin{eqnarray}
  & e^+ + e^- \rightarrow \gamma + \nu_l + \bar{\nu}_l, \quad l = e, \mu, \tau. \label{background}
\end{eqnarray}
To extract the important information from the signal at a high confidence level, it is necessary to reduce the background, and hence enhancing the signal-to-noise ratio. We note that the neutrinos in the SM are left-handed particles. So the t- and u-channels of Eq. (\ref{background}) with the W-boson exchange are suppressed by using the right-handed electron beam. In future linear colliders, it is possible to use both polarized initial beams enabling us to suppress the background even more. The cross-section of the scattering process involving both the partially polarized initial beams can be determined as follows:
\begin{equation}
\begin{array}{llc}
\sigma(e^+e^-) & = & (1 - p^+)(1 - p^-)\sigma_{LL} + (1 - p^+)p^-\sigma_{LR} \\
               &   & + p^+(1 - p^-)\sigma_{RL} + p^+p^-\sigma_{RR},
\end{array}
\label{cross-section}
\end{equation}
where $p^+$, $p^-$ are the right-handed polarization degrees of the positron and electron beams, $\sigma_{LL}$, $\sigma_{LR}$, $\sigma_{RL}$ and $\sigma_{RR}$ are the cross-sections of the fully polarized incoming beams $e^+_L e^-_L$, $e^+_L e^-_R$, $e^+_R e^-_L$, and $e^+_R e^-_R$ respectively.
According to the recent achievement in producing polarized electron and positron beams (Refs. \refcite{polarization1,polarization2}), in our calculation, we assume the 80\% left-handed positron beam and the 90\% right-handed electron beam at the future $e^+e^-$ collision which will be shown in the next section to be the best choice of polarization combination.

It is also essential to note that the region around the Z-resonance peak of the photon energy distribution of the background cross-section contributes much to the total cross-section. For the collision with $\sqrt{s} = 1$ TeV, the center of this peak is at the value of photon energy:
\begin{eqnarray}
E_{\gamma}^{(Z)} = \frac{s - m_Z^2}{2 \sqrt{s}} \approx 496 \; \rm GeV.
\end{eqnarray}
Besides, the photon trigger only triggers events when the energy amount in the calorimeter goes beyond a certain threshold. So in our consideration, we apply the following cuts on the photon energy:
\begin{eqnarray}
10 \; \rm{GeV} \leq E_{\gamma} \leq 400 \; \rm{GeV}
\end{eqnarray}
to cut away the large contribution due to the Z on-shell exchange region via the s-channel, while the SUSY signal is still almost the same because there is no Z-resonance in the photon energy distribution of the signal cross-section in the scenarios with $m_{\tilde{\chi}^0_1}, \; m_{\tilde{\nu}_l} > m_Z/2$. The minimum energy cut helps to regularize the infrared divergences of the tree level cross-sections.

Another point is that, because of the beam pipe, the detectors cannot cover the whole polar angle leading to some missing amount of single-photon events. This fact is taken into account by using the cuts on the photon polar angle:
\begin{eqnarray}
10^\circ \leq \theta_\gamma \leq 170^\circ.
\end{eqnarray}
The collinear divergences are also regularized thanks to these cuts.

In this paper, the luminosity $L = 1000 \; \rm fb^{-1}/year$ is expected at the future $e^+e^-$ collision and we estimate how long it will take to see the signal difference between the two models exceeding three times the statistical error. To show how significant the signal is, beside the signal-to-noise ratio:
\begin{eqnarray}
R = \dfrac{N_S}{N_B},
\end{eqnarray}
we also calculate the statistical significance defined as:
\begin{eqnarray}
S = \dfrac{N_S}{\sqrt{N_S + N_B}},
\end{eqnarray}
where $N_S$ and $N_B$ are  respectively the numbers of events for the signal and background processes after a given duration of data accumulation.

\section{Results}

\begin{figure}
\begin{center}
\includegraphics[scale=0.49]{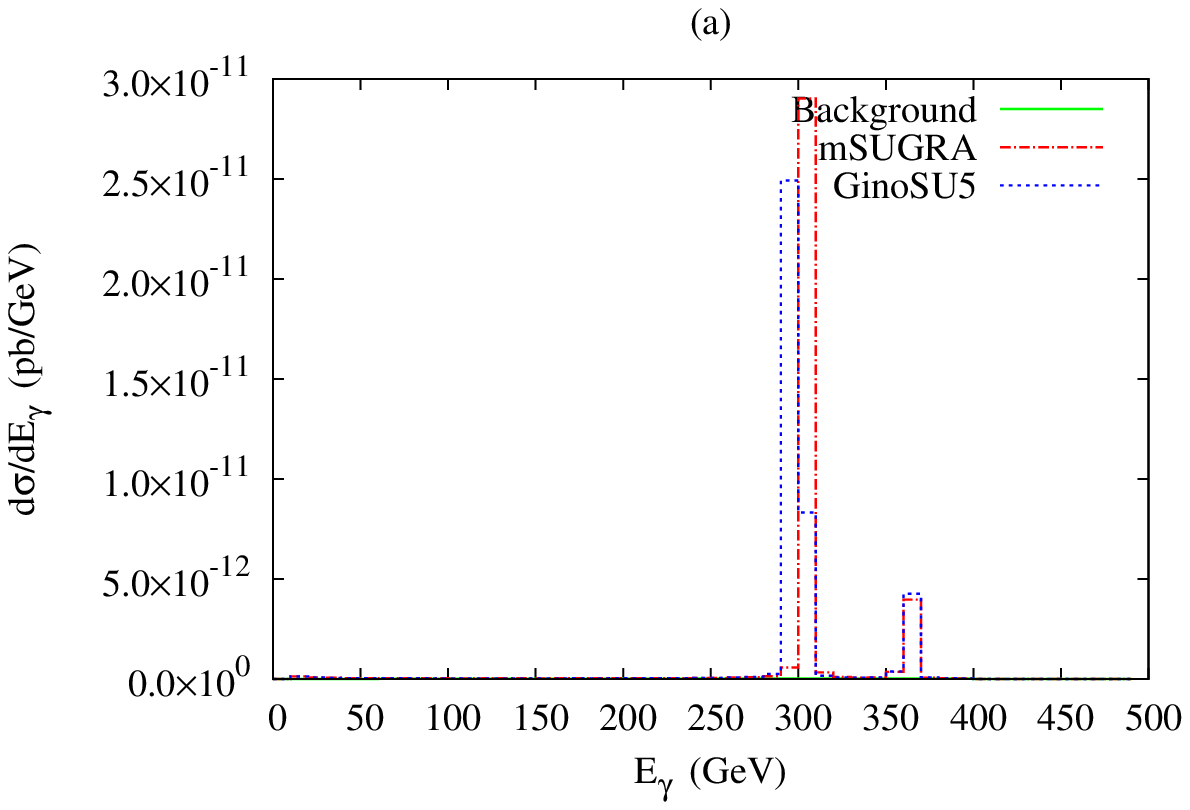}
\includegraphics[scale=0.49]{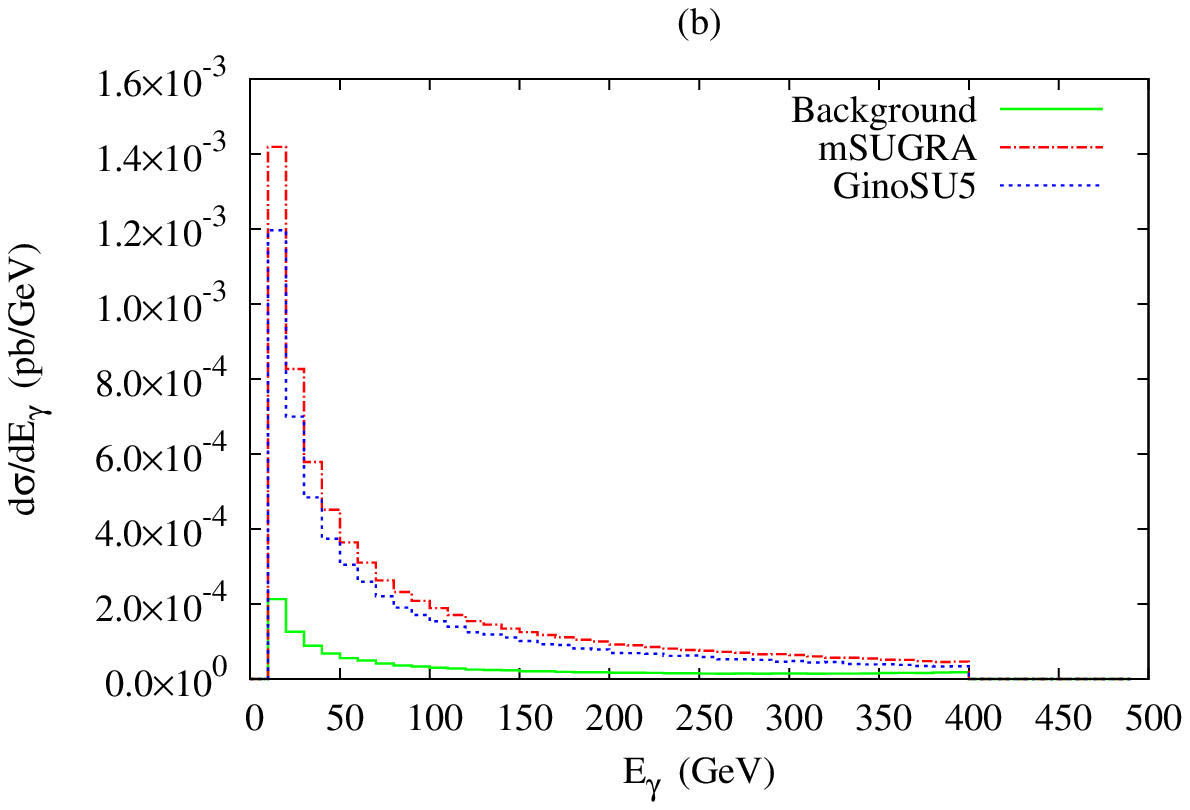}
\includegraphics[scale=0.49]{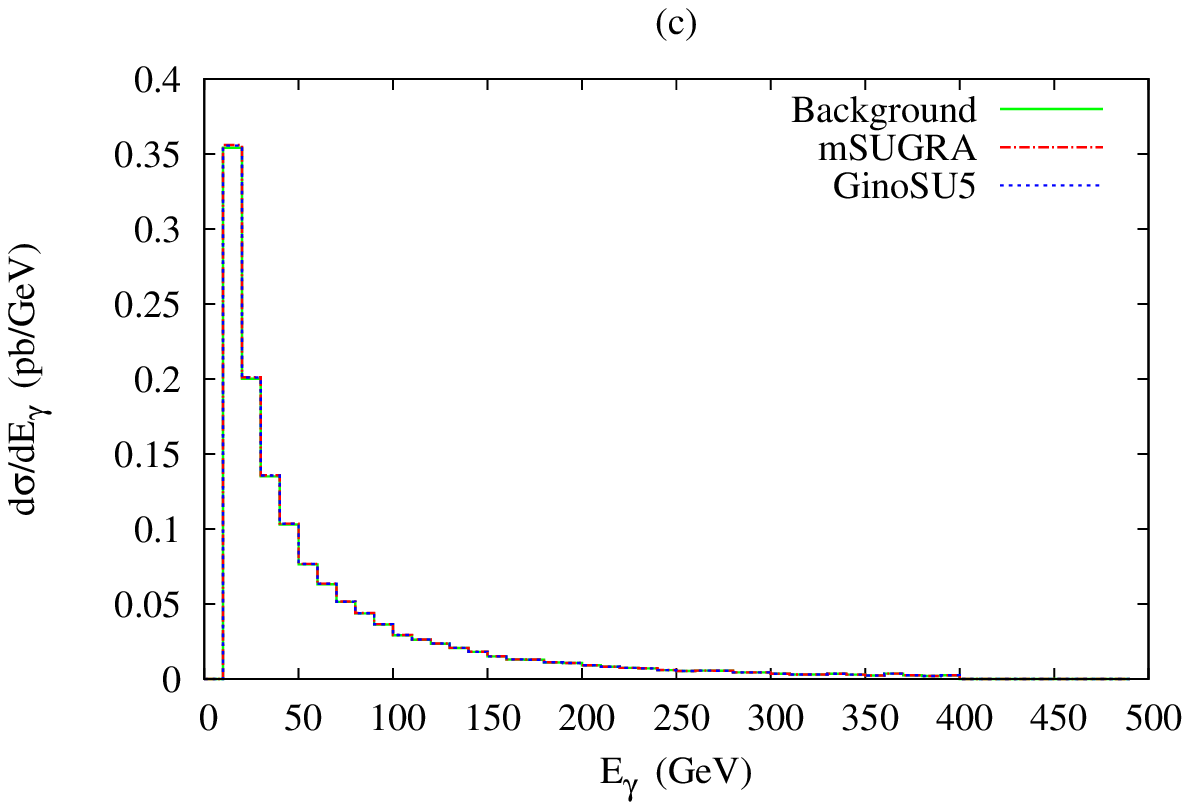}
\includegraphics[scale=0.49]{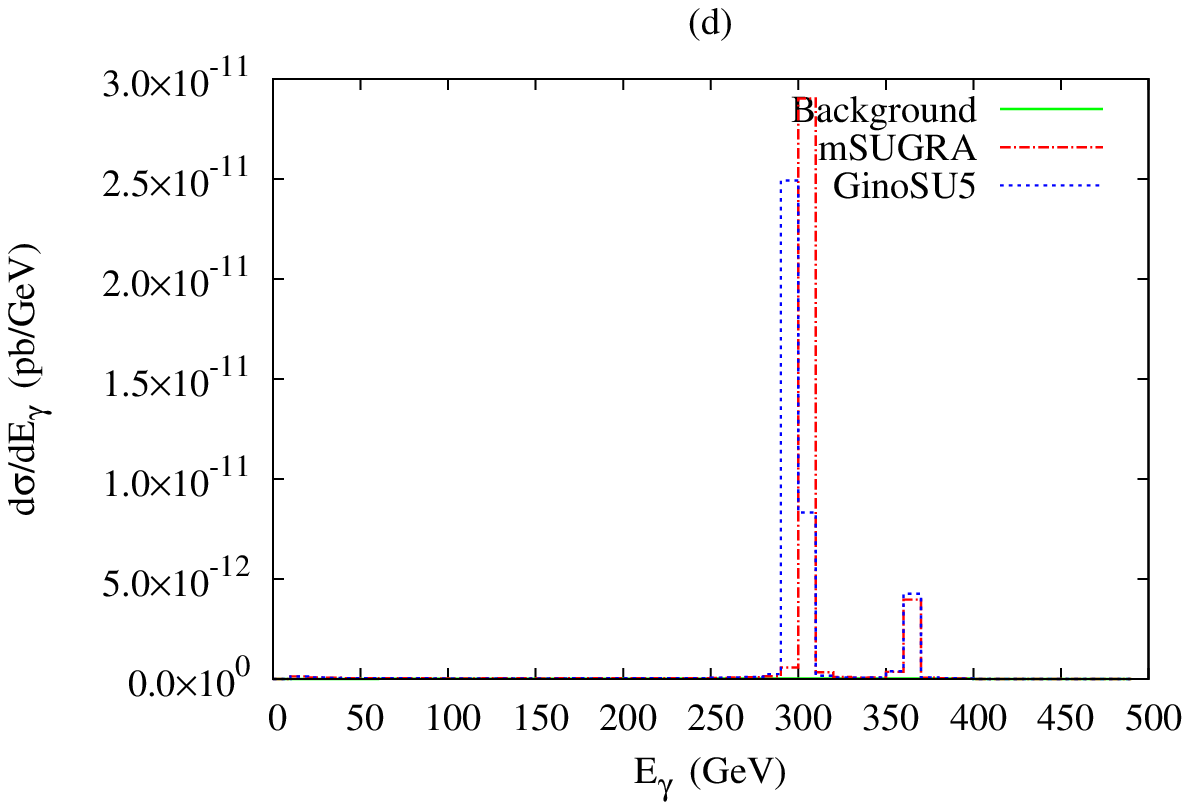}
\caption{Photon energy distributions of the single-photon cross-sections for all the possible polarization combinations: (a) $e^+_L e^-_L$, (b) $e^+_L e^-_R$, (c) $e^+_R e^-_L$ and (d) $e^+_R e^-_R$. While the solid (green) lines indicate the SM background distributions, the dot-dashed (red) and dotted (blue) lines correspond to the sum of both signal and background distributions in the mSUGRA and GinoSU5 models.}
\label{energy_distributions}
\end{center}
\end{figure}

\begin{figure}
\begin{center}
\includegraphics[scale=0.49]{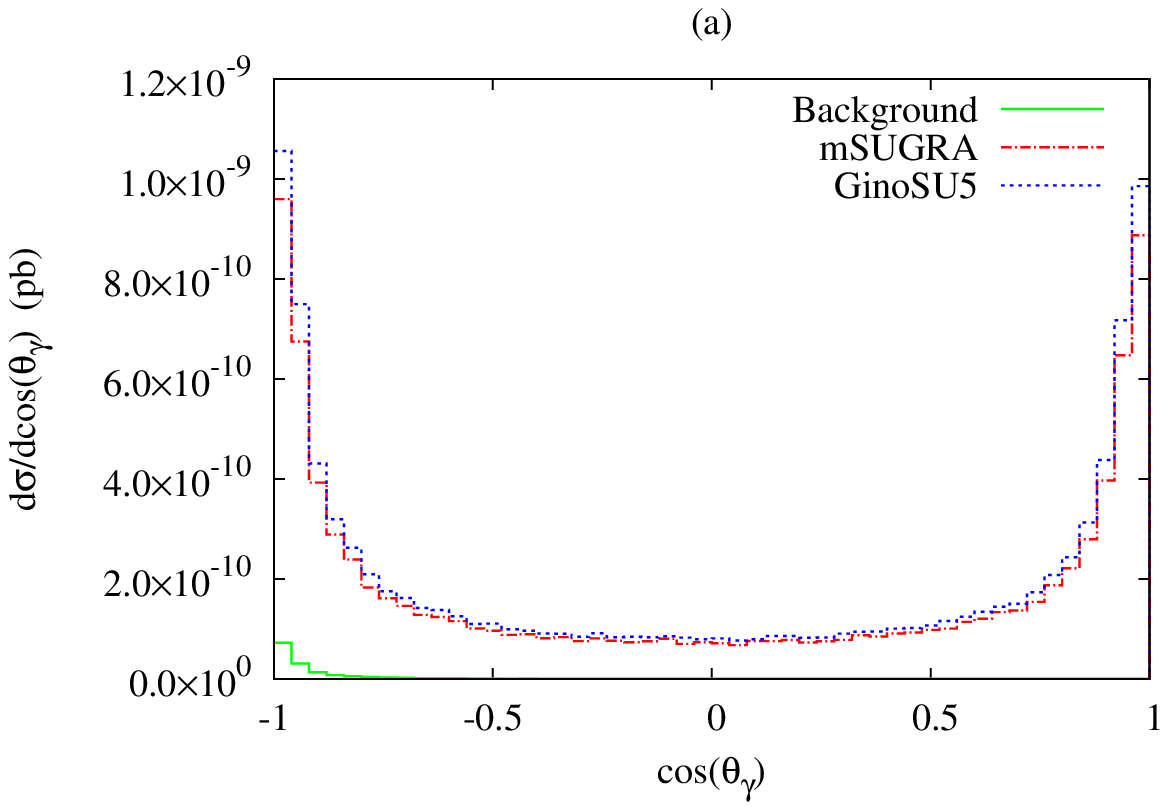}
\includegraphics[scale=0.49]{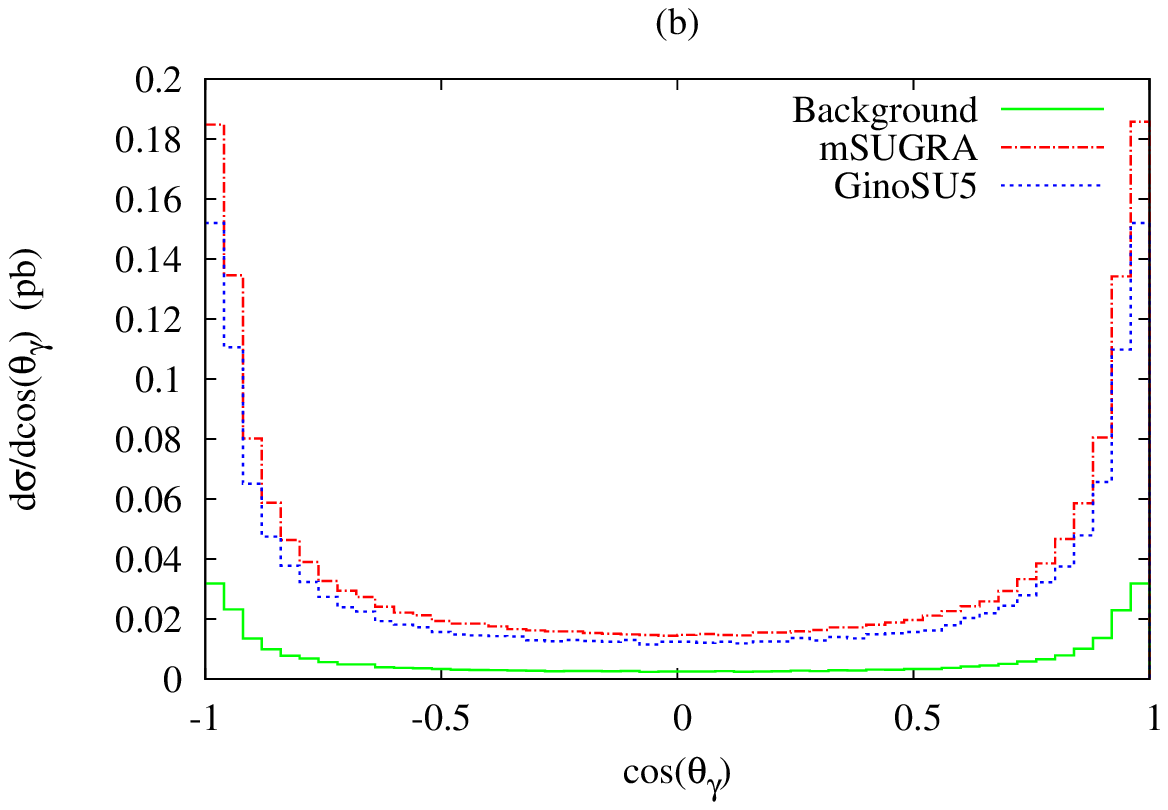}
\includegraphics[scale=0.49]{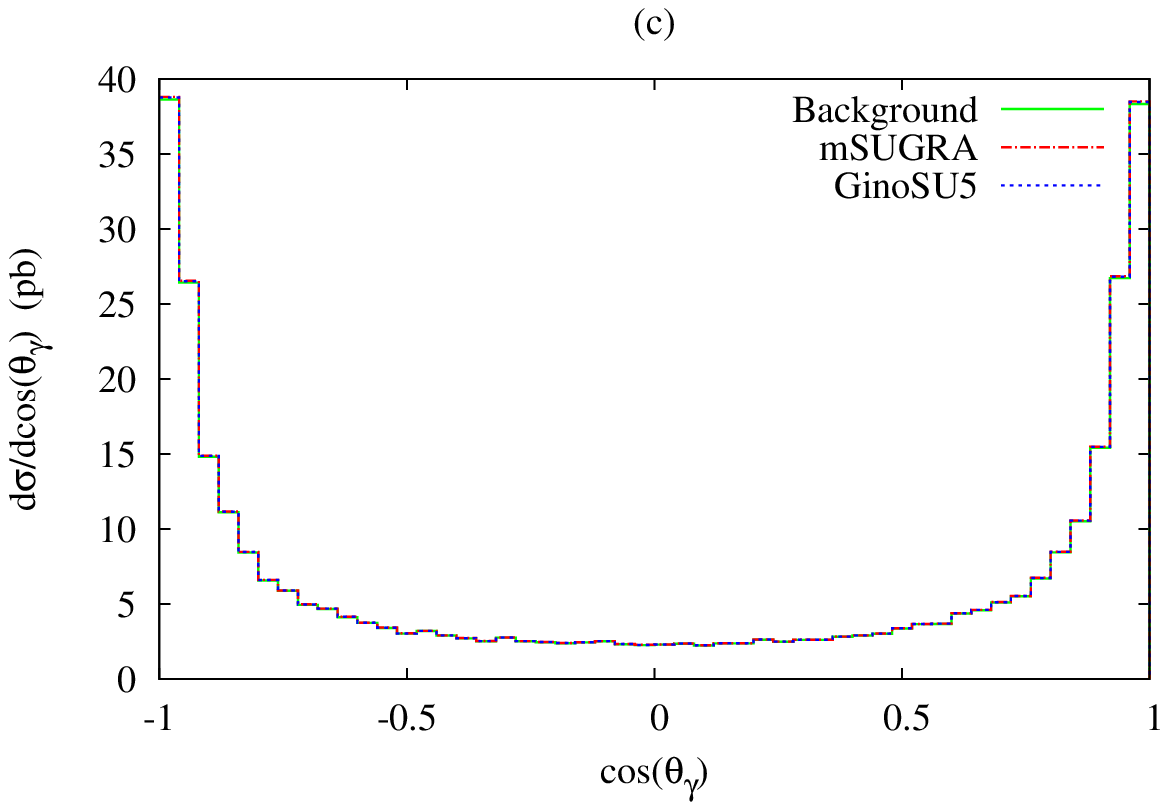}
\includegraphics[scale=0.49]{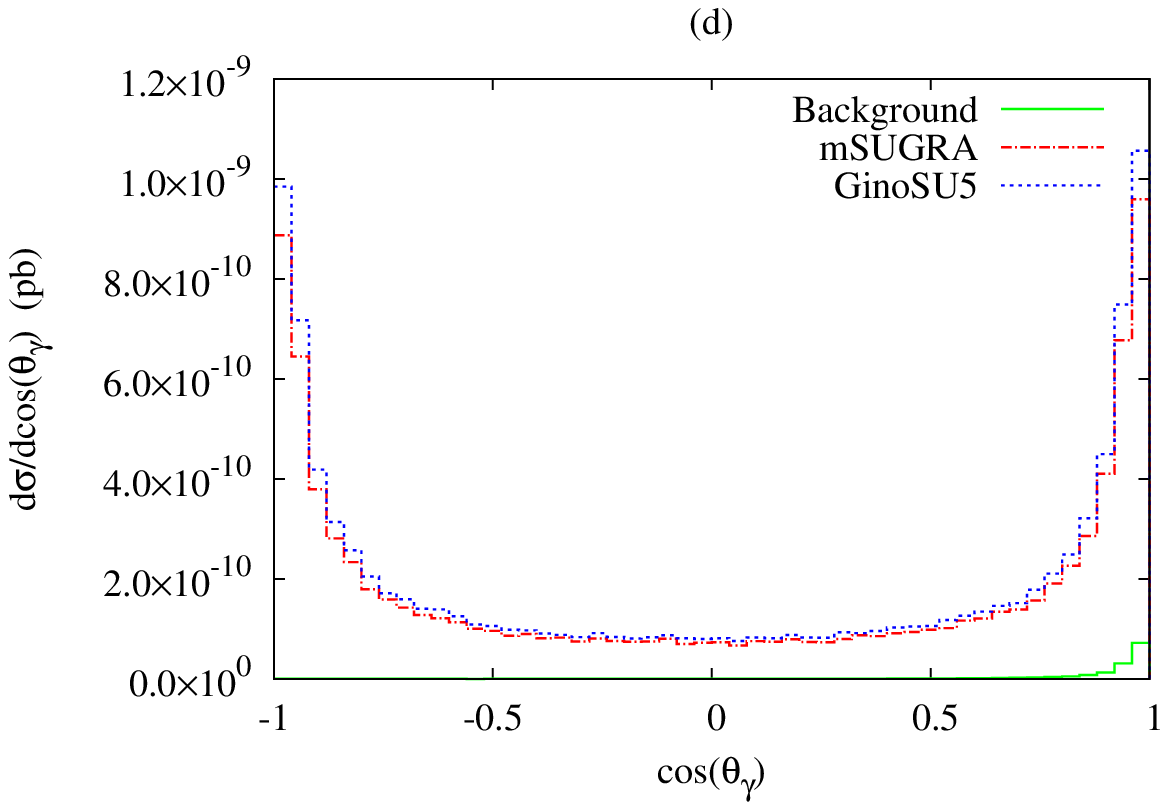}
\caption{$ \cos (\theta_{\gamma}) $ distributions of the single-photon cross-sections for all the possible polarization combinations: (a) $e^+_L e^-_L$, (b) $e^+_L e^-_R$, (c) $e^+_R e^-_L$ and (d) $e^+_R e^-_R$. The line conventions in the caption of Fig. \ref{energy_distributions} are still used in this figure.}
\label{theta_distributions}
\end{center}
\end{figure}

Figure \ref{energy_distributions} shows the photon energy distributions of the cross-sections corresponding to all the possible polarization combinations of the initial positron and electron beams. The cross-sections with $e^+_L e^-_L$ (Fig. \ref{energy_distributions}a) and $e^+_R e^-_R$ (Fig. \ref{energy_distributions}d) are extremely suppressed by the beam polarization. We only see the remaining peaks due to the heavier CP-even Higgs and CP-odd Higgs resonance exchanges through the s-channel. In Figs. \ref{energy_distributions}b and \ref{energy_distributions}c, we see that the most important contributions to these distributions come from the low photon energy region.
Similar to Fig. \ref{energy_distributions}, in Fig. \ref{theta_distributions} the photon polar angle distributions of the cross-sections for all the polarization combinations of the incoming beams are plotted. From this figure, it is obvious that the distributions are dominated by the events with their photons going close to the beam line direction. The forward-backward asymmetry relevant to the background processes is observed in Figs. \ref{theta_distributions}a and \ref{theta_distributions}d, while such asymmetry is not clear in Figs. \ref{theta_distributions}b and \ref{theta_distributions}c.

The cross-sections of the background and signal processes relevant to the single-photon events corresponding to all the polarization combinations are summarized in Table \ref{result}. Here the decay widths and the branching ratios of the visible and invisible decay channels of the scalar neutrinos are also presented. Due to the extremely small cross-sections, the interactions between $e^+_L$ and $e^-_L$, $e^+_R$ and $e^-_R$ are negligible. The remaining important polarization combinations are $e^+_L e^-_R$ and $e^+_R e^-_L$. In the $e^+_R e^-_L$ collision, the SM background is about three orders of magnitude larger than the SUSY signal giving a very small signal-to-noise ratio. While in the $e^+_L e^-_R$ collision, the background is suppressed such that it is even smaller than the signal providing the ability to discriminate between the SUSY models.

\begin{figure}
\begin{center}
\includegraphics[scale=0.49]{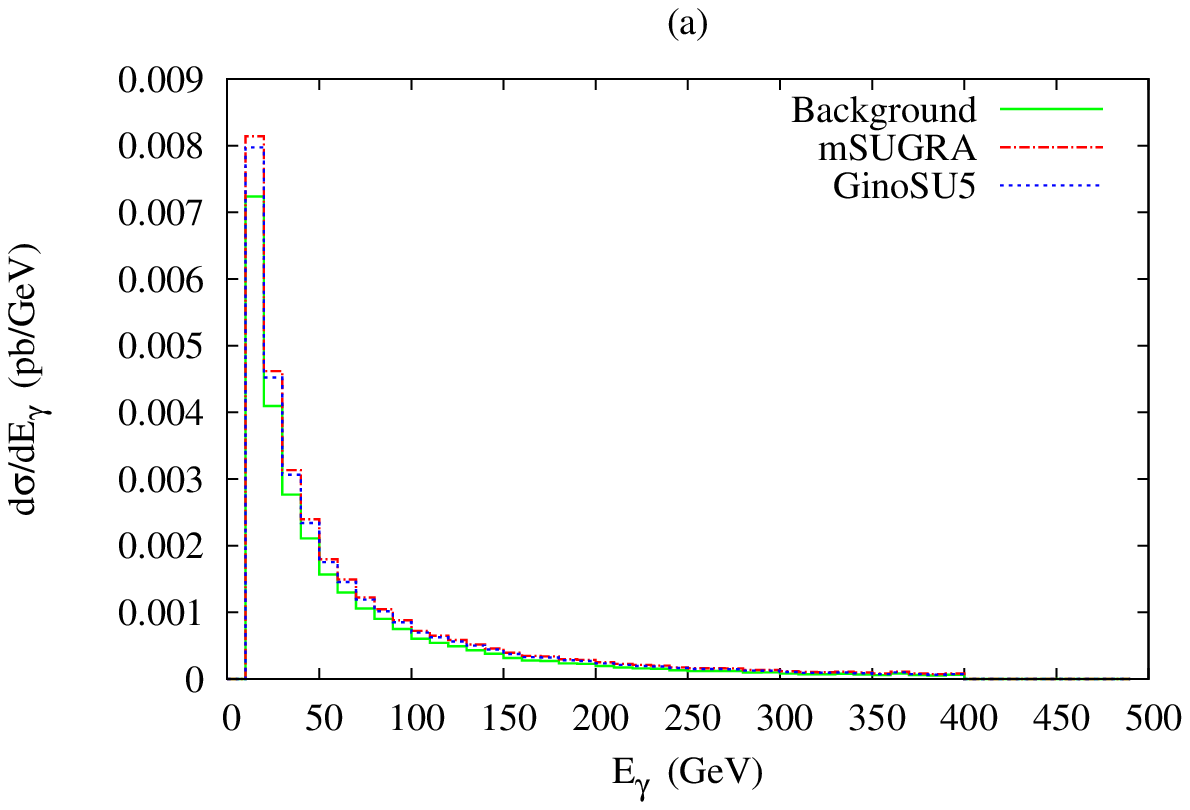}
\includegraphics[scale=0.49]{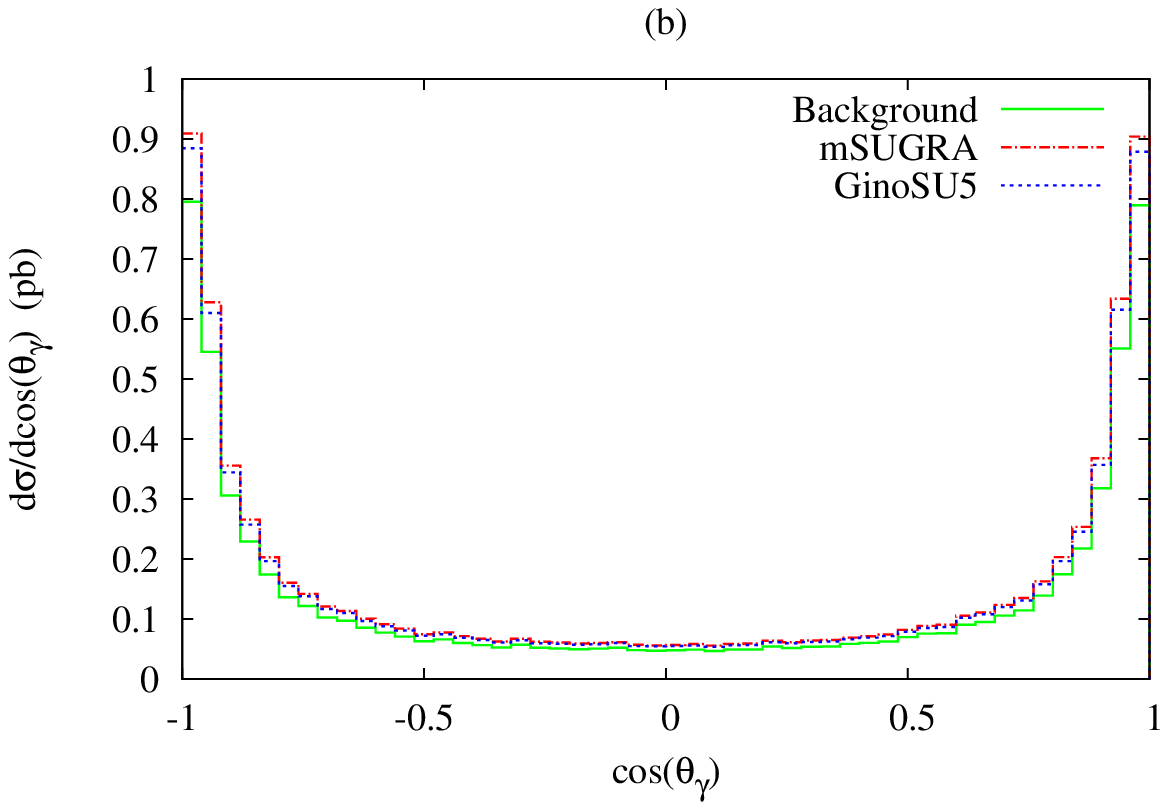}
\caption{Distributions of cross-sections in the case of $80\%$ left-handed $e^{+}$ beam and $90\%$ right-handed $e^{-}$ beam with respect to: (a)  photon energy, and (b) photon polar angle. The line conventions in the caption of Fig. \ref{energy_distributions} are still used in this figure.}
\label{distributions}
\end{center}
\end{figure}

Since in practice, it is impossible to produce purely polarized beams, we assume in the future running of the ILC the 80\% left-handed positron beam and the 90\% right-handed electron beam which have been recently achieved. The differential cross-sections with respect to the photon$'$s energy and polar angle in the collision of the above partly polarized beams are plotted in Fig. \ref{distributions}. Using Eq. (\ref{cross-section}), we obtain the following results: the background cross-section is 0.276 pb, the signal cross-sections of the mSUGA and GinoSU5 models are 0.045 pb and 0.035 pb respectively. The mSUGRA signal is larger than the GinoSU5 one because the slepton masses in the former model are lighter than those in the latter one. As the consequence, the signal-to-noise ratios for the two models are: $R_{\rm mSUGRA} = 16.3 \%$ and $R_{\rm GinoSU5} = 12.7 \%$. With the luminosity $L =1000 \; \rm fb^{-1}/year$, we find that it requires at least three years of data accumulating to clearly see the difference between the two models, namely the signal difference would exceed three times the statistical error. After three years of running, the expected numbers of events for the background and signal processes of the two models are respectively: $N_B = 8277$, $N_S^{\rm mSUGRA} = 1352$, $N_S^{\rm GinoSU5} = 1054$. Hence the statistical significances are: $S_{\rm mSUGRA} = 13.8$, $S_{\rm GinoSU5} = 10.9$. These results give us the possibility to probe the SUSY breaking models using the single-photon events at future $e^+ e^-$ linear colliders, especially the ILC.

\section{Summary and discussions}

We have considered in this paper the single-photon signal in future $e^+ e^-$ linear colliders and found that it is possible to probe SUSY breaking models using this kind of signal. The mSUGRA and GinoSU5 models have been taken into account as examples. Starting from the given benchmark points of the free parameter spaces of the two models which produce a common neutralino-LSP mass and satisfy various phenomenological constraints, we then obtained the mass spectra  and the mixing angles by solving the RG equations. Subsequently, the cross-sections of the single-photon processes were computed. After three years of data accumulation, the difference between the two models would be large enough to see which one is realized in nature. These results also tell us that the single-photon data collected from future $e^+ e^-$ colliders can be used to build up an independent constraint on SUSY breaking models.

It has been previously proofed that the full one-loop electroweak radiative corrections to the single-photon background processes in the $e^+ e^-$ collision at $\sqrt{s} = 1$ TeV is about 1\% of the tree-level cross-section.\cite{1-loop_bg} This amount is negligible in our analysis since it adds only a tiny part to the statistical error. After assuming a common base and constraints for the SUSY breaking models, the difference between mass spectra is not so large, namely the sparticle masses of the same type are of the same order. Thus the SUSY loop-corrections would enter almost the same amounts to the cross-sections at the Born approximation in the two scenarios. It follows that the signal difference between the two models does not change significantly, while only the extra number of events of one model from the other is crucial to distinguish between models.

Since we are dealing with the single-photon events at $e^+ e^-$ colliders, the dominant contributions to the SUSY signal difference come from the neutralino and slepton sectors (the Higgs sector does not give important contributions to signal due to the small Yukawa couplings of the first generation). If we take into account the cosmological constraint on the dark matter relic density, the mass difference between the two models will be very small because of the neutralino-stau coannihilation condition. The expected mass difference can be found in the right-handed down-type squark and left-handed slepton sectors. Therefore, in this case, it requires extremely high polarization degrees to suppress the background more, longer time of data accumulation to discriminate between SUSY breaking models using this kind of events. Our analysis still holds in the frame work of non-standard cosmology where the dark matter constraint can be relaxed.\cite{non-standard_cosmology1}$^-$\cite{non-standard_cosmology5}

\section*{Acknowledgement}

H.M.T. would like to thank the Organizers of KEK-Vietnam Visiting Program for hospitality and support during his visit. He is also grateful to Nobuchika Okada for useful discussions and comments.

\begin{landscape}
\begin{table}
\tbl{Signal and background of single-photon processes for all possible polarization combinations. Cross-sections are in pb, and decay widths are in GeV. Bold numbers are the total cross-sections of background or signal processes. ``NO'' in some decay channels indicate that they are kinematically forbidden.}
{\begin{math}
\begin{array}{|c|c|c|c|c|c|c|c|c|c|}
\hline
\multicolumn{2}{|c|}{\multirow{2}*{\bf Polarization}} & \multicolumn{3}{|c|}{e^+} & {\bf L} & {\bf L} & {\bf R} & {\bf R} & \multirow{2}*{{\bf Branching ratio}} \\

  \multicolumn{2}{|c|}{}  & \multicolumn{3}{|c|}{e^-} & {\bf L} &  {\bf R} &  {\bf L} &  {\bf R} & \\
\hline

\multicolumn{2}{|c|}{\multirow{4}*{\bf Background}} & \multicolumn{3}{|c|}{\nu_e} & 6.4335 \times 10^{-12} & 4.1421 \times 10^{-3} & 1.3335 \times 10^{01} & 6.3901 \times 10^{-12} & \\

 \multicolumn{2}{|c|}{} & \multicolumn{3}{|c|}{\nu_\mu} & 4.2671 \times 10^{-15} & 4.1421 \times 10^{-3} & 6.0749 \times 10^{-3} & 4.2617 \times 10^{-15} &  \\

 \multicolumn{2}{|c|}{} & \multicolumn{3}{|c|}{\nu_\tau} & 4.2671 \times 10^{-15} & 4.1421 \times 10^{-3} & 6.0749 \times 10^{-3} & 4.2617 \times 10^{-15} &  \\
\cline{3-5}

 \multicolumn{2}{|c|}{} & \multicolumn{3}{|c|}{{\bf total}} & \bf{6.4421 \times 10^{-12}} & \bf{1.2426 \times 10^{-2}} & \bf{1.3347 \times 10^{01}} & \bf{6.3986 \times 10^{-12}} & \\
\hline

\multirow{22}*{\bf Signal} & \multicolumn{1}{|c|}{\multirow{11}*{\bf mSUGRA}} & \multicolumn{3}{|c|}{\chi^0_1} & 3.4929 \times 10^{-10} & 5.8297 \times 10^{-2} & 2.2217 \times 10^{-3} & 3.4925 \times 10^{-10} & \\
\cline{3-5}

 &  & \multicolumn{1}{|c|}{\multirow{3}*{$\tilde{\nu}_e$}} & \multicolumn{2}{|c|}{\rm production} & 3.8356 \times 10^{-12} & 7.8733 \times 10^{-4} & 6.5127 \times 10^{-2} & 3.8338 \times 10^{-12} & \\
\cline{4-10}

 &  &  & \multicolumn{1}{|c|}{\multirow{2}*{decay}} & \rm{invisible \; channel} & \multicolumn{4}{|c|}{1.6491 \times 10^{-1}} & 1.0000 \\

 &  &  &  & \rm{visible \; channel} & \multicolumn{4}{|c|}{\rm NO} & 0.0000 \\
\cline{3-10}

 &  & \multicolumn{1}{|c|}{\multirow{3}*{$\tilde{\nu}_\mu$}} & \multicolumn{2}{|c|}{\rm production} & 2.2840 \times 10^{-14} & 7.8733 \times 10^{-4} & 1.1559 \times 10^{-3} & 2.2846 \times 10^{-14} & \\
\cline{4-10}

 &  &  & \multicolumn{1}{|c|}{\multirow{2}*{decay}} & \rm{invisible \; channel} & \multicolumn{4}{|c|}{1.6491 \times 10^{-1}} & 1.0000 \\

 &  &  &  & \rm{visible \; channel} & \multicolumn{4}{|c|}{\rm NO} & 0.0000 \\
\cline{3-10}

 &  & \multicolumn{1}{|c|}{\multirow{3}*{$\tilde{\nu}_\tau$}} & \multicolumn{2}{|c|}{\rm production} & 2.3433 \times 10^{-14} & 7.9384 \times 10^{-4} & 1.1655 \times 10^{-3} & 2.3439 \times 10^{-14} & \\
\cline{4-10}

 &  &  & \multicolumn{1}{|c|}{\multirow{2}*{decay}} & \rm{invisible \; channel} & \multicolumn{4}{|c|}{1.6298 \times 10^{-001}} & 1.0000 \\

 &  &  &  & \rm{visible \; channel} & \multicolumn{4}{|c|}{\rm NO} & 0.0000 \\
\cline{3-10}

 &  & \multicolumn{3}{|c|}{\bf total} & \bf{3.5318 \times 10^{-10}} & \bf{6.0665 \times 10^{-2}} & \bf{6.9670 \times 10^{-2}} & \bf{3.5313 \times 10^{-10}} &  \\
\cline{2-10}

 & \multicolumn{1}{|c|}{\multirow{11}*{\bf GinoSU5}} & \multicolumn{3}{|c|}{\tilde{\chi^0_1}} & 3.8862 \times 10^{-10} & 4.5698 \times 10^{-2} & 1.9769 \times 10^{-3} & 3.8868 \times 10^{-10} &  \\
\cline{3-5}

 &  & \multicolumn{1}{|c|}{\multirow{3}*{$\tilde{\nu}_e$}} & \multicolumn{2}{|c|}{\rm production} & 3.2740 \times 10^{-12} & 5.6117 \times 10^{-4} & 4.9549 \times 10^{-2} & 3.2728 \times 10^{-12} &  \\
\cline{4-10}

 &  &  & \multicolumn{1}{|c|}{\multirow{2}*{decay}} & \rm{invisible \; channel} & \multicolumn{4}{|c|}{2.3542 \times 10^{-1}} & 9.7172 \times 10^{-1} \\

 &  &  &  & \rm{visible \; channel} & \multicolumn{4}{|c|}{6.8520 \times 10^{-3}} & 2.8282 \times 10^{-2} \\
\cline{3-10}

 &  & \multicolumn{1}{|c|}{\multirow{3}*{$\tilde{\nu}_\mu$}} & \multicolumn{2}{|c|}{\rm production} & 1.2439 \times 10^{-14} & 5.6117 \times 10^{-4} & 8.2387 \times 10^{-4} & 1.2450 \times 10^{-14} & \\
\cline{4-10}

 &  &  & \multicolumn{1}{|c|}{\multirow{2}*{decay}} & \rm{invisible \; channel} & \multicolumn{4}{|c|}{2.3542 \times 10^{-1}} & 9.7172 \times 10^{-1} \\

 &  &  &  & \rm{visible \; channel} & \multicolumn{4}{|c|}{6.8509 \times 10^{-3}} & 2.8278 \times 10^{-2} \\
\cline{3-10}

 &  & \multicolumn{1}{|c|}{\multirow{3}*{$\tilde{\nu}_\tau$}} & \multicolumn{2}{|c|}{\rm production} & 1.2594 \times 10^{-14} & 5.6705 \times 10^{-4} & 8.3251 \times 10^{-4} & 1.2605 \times 10^{-14} &  \\
\cline{4-10}

 &  &  & \multicolumn{1}{|c|}{\multirow{2}*{decay}} & \rm{invisible \; channel} & \multicolumn{ 4}{|c|}{2.3353 \times 10^{-1}} & 9.7905 \times 10^{-1} \\

 &  &  &  & \rm{visible \; channel} & \multicolumn{4}{|c|}{4.9969 \times 10^{-3}} & 2.0949 \times 10^{-2} \\
\cline{3-10}

 &  & \multicolumn{3}{|c|}{{\bf total}} & \bf{3.9183 \times 10^{-10}} & \bf{4.7344 \times 10^{-2}} & \bf{5.1740 \times 10^{-2}} & \bf{3.9189 \times 10^{-10}} & \\
\hline
\end{array}
\end{math} }
\label{result}
\end{table}
\end{landscape}

\end{document}